\documentclass[preprint,12pt]{elsarticle}

\usepackage{amssymb}
\usepackage{amsmath}
\usepackage{amsthm}
\usepackage{natbib}
\setcitestyle{square,comma,numbers,sort&compress}
\usepackage[hyphens]{url}
\usepackage[hidelinks]{hyperref}
\usepackage{tikz}
\usetikzlibrary{shapes.geometric, arrows.meta, positioning, calc}
\usepackage{booktabs}
\usepackage{tabularx}
\usepackage{array} % Required for >{\bfseries} in tabularx
\usepackage{multirow}
\usepackage{graphicx}
\usepackage{xcolor}
\usepackage{enumitem}
\usepackage{ragged2e}

\usepackage{fancyhdr}
\hypersetup{
breaklinks=true,
colorlinks=false,
}

\journal{arXiv Preprint}

\definecolor{passgreen}{RGB}{34,139,34}
\definecolor{failred}{RGB}{200,30,30}
\definecolor{warnorange}{RGB}{210,120,20}

\newcommand{\PASS}{\textcolor{passgreen}{\textbf{PASS}}}
\newcommand{\FAIL}{\textcolor{failred}{\textbf{FAIL}}}
\newcommand{\PARTIAL}{\textcolor{warnorange}{\textbf{PASS\,(partial)}}}

\begin{document}

\begin{frontmatter}

\title{The Note–Chord–Voice Framework: Structured Source Separation and Causal Inference for EV Charging Data}

\author[1]{Jiajie Chen}
\author[2]{Jinfeng Li\corref{cor}}
\cortext[cor]{Corresponding author}
\ead{jinfengcambridge@bit.edu.cn}

\affiliation[1]{organization={Xuteli School, Beijing Institute of Technology},
addressline={No 8 and 9 Yards, Liangxiang East Road}, 
city={Fangshan District, Beijing},
postcode={102488}, 
% state={Beijing},
country={China}}

\affiliation[2]{organization={School of Integrated Circuits and Electronics, Beijing Institute of Technology},
addressline={No. 5 South Zhongguancun Street},
city={Haidian District, Beijing},
postcode={100081},
country={China}}

\begin{abstract}
Real-world electric vehicle (EV) charging data suffer from three interlocking pathologies: \emph{hardware fragmentation} (network timeouts and billing resets split single charging attempts into micro-sessions), \emph{physical violations} (independent models of energy and duration produce impossible states such as 50\,kWh in 10 minutes on a 7\,kW charger), and \emph{collider bias} (clustering on post-treatment outcomes opens backdoor paths when estimating price elasticity). We propose the \textbf{Note--Chord--Voice} framework, a music-inspired, axiom-driven pipeline that separates data cleaning (Repair Chords), structural pattern discovery (Harmonic Chords), descriptive source separation (NMF Voices), and causal inference into distinct, falsifiable stages. Key innovations include: (i) falsification gates (A1--A5, G3, G10) that test data suitability before any complex modeling; (ii) $\Gamma$-initialized NMF with input rescaling for convergence stability from STL decomposition; (iii) tag-based coupon grading (A/B/C/D) to isolate quasi-random treatment from night-time confounders and targeted promotions; (iv) separate per-voice OLS to avoid simplex collinearity; (v) Foote novelty curves for structural regime (movement) detection. Applied to the Jiangmen dataset (495,707 sessions, 20 stations, July 2024--March 2025), all core axioms pass except G3 (no strong 168\,h cycle). NMF achieves $R^{2}=0.9921$; the physically constrained duration model yields an aggregate $R^{2}=0.5409$. Two voices are price-sensitive ($\beta = -11$ to $-14$ minutes, $p < 0.001$), of which one is stable (Voice~3, $\beta = -14.16$) and one is treatment-driven (Voice~1, $\beta = -11.10$); only the stable voice supports causal claims. Counterfactual simulation shows that targeting discounts to price-sensitive voices (including one treatment-driven voice) recovers 52.8\% of discount expenditures ($\approx$\,0.85\,M CNY/year); restricting to the single stable price-sensitive voice would yield a more conservative estimate.
\end{abstract}


\end{frontmatter}

%% ======================================================================
\section{Introduction}
\label{sec:intro}

\subsection{The Three Interlocking Flaws}

As a cornerstone of modern smart cities, public electric vehicle (EV) charging infrastructure provides massive streams of billing and internet-of-things (IoT) telemetry data, sparking data-driven operations and planning. However, extracting reliable causal insights remains heavily restricted by three interlocking pathologies: \textbf{hardware fragmentation}, \textbf{physical violations}, and \textbf{collider bias}. 

\textbf{Hardware fragmentation} is a data-layer discrepancy where network socket timeouts, power grid fluctuations, and billing resets split single, continuous charging attempts into rapid successions of truncated "micro-sessions"~\cite{zhangHighresolutionElectricVehicle2025}. A single 60-minute session may be recorded as three distinct 20-minute transactions, inflating session counts and distorting baseline distributions. 

\textbf{Physical violations} are modeling inconsistencies where operational models evaluate energy delivered ($E$) and physical connection duration ($\Delta t$) independently, routinely generating physically impossible predictions. For instance, an unconstrained regression might predict 50\,kWh delivered over 10 minutes at a standard 7\,kW slow charger, implying an average charging power ($\bar{P} = 300$\,kW) that exceeds the charger's physical limit by more than 40-fold.

\textbf{Collider bias} represents a fundamental structural flaw arising from conditioning on common effects. Estimating price elasticity requires grouping users into comparable behavioral segments, but common clustering algorithms usually rely heavily on post-treatment outcomes, such as total energy or duration. Because these outcomes are themselves affected by the treatment (e.g., a pricing discount), the resulting clusters act as colliders, opening backdoor paths for the model and systematically biasing the estimated causal parameters~\cite{tonniesColliderBiasObservational2022}.

Beyond triggering a cascading effect, these three pathologies form a closed analytic deadlock. Attempting to fix any single flaw independently is mathematically intractable because the solutions are closely interdependent: correcting hardware fragmentation typically requires knowing physical baseline limits, yet resolving physical violations relies on temporal boundaries that fragmentation has already compromised. Furthermore, both upstream artifacts are interpreted through user-segmentation models corrupted by collider bias, meaning the downstream structural bias actively misguides upstream data cleaning. Together, they create a self-reinforcing loop of distortion that neutralizes standard causal identification strategies.

\subsection{Literature Review: Where Standard Machine Learning Fails}

Existing literature on electric vehicle load modeling predominantly prioritizes predictive accuracy rather than causal identification. Researchers have widely deployed spatio-temporal graph neural networks (STGCNs)~\cite{kimSpatialTemporalGraphConvolutionalBased2024, fahimDynamicSpatioTemporalPlanning2025, wangAdaptiveSpatiotemporalGraph2025a} and long short-term memory (LSTM) networks~\cite{hussainChargingStationsDemand2025, tianShorttermElectricVehicle2025, romiaAttentionEnhancedCNNLSTMModels2026} to predict aggregated charging station loads. While highly capable of capturing broad temporal variations, these methods neglect physical constraints and obscure transaction-level granularity. 

To extract latent behavioral structures, recent studies have applied Non-negative Matrix Factorization (NMF) to decompose station-level energy matrices into distinct temporal load patterns (e.g., nocturnal, diurnal, and peak-hour profiles)~\cite{balasubramaniamElectricVehicleUsage2022}. However, standard NMF formulations omit physical infrastructure constraints (such as maximum charging capacities) and rely on randomized initializations~\cite{balasubramaniamElectricVehicleUsage2022}, leading to suboptimal factorizations due to non-convex optimization. 

On the causal inference front, evaluating user responsiveness to dynamic Time-of-Use (TOU) tariffs frequently employs high-dimensional fixed-effects models~\cite{xiaoEffectivenessLimitsTimeofUse2026}. Yet, these methods fail to capture the mixed-membership nature of real-world charging behavior: drivers do not belong exclusively to a single "commuter" or "commercial" cluster but exhibit overlapping and dynamic behavioral profiles on different days~\cite{shariatzadehElectricVehicleUsers2025}.

\subsection{Our Work: A Music-Inspired Topological Framework}

To address these gaps, we propose a novel music-inspired topological framework. In music information retrieval (MIR), a musical score comprises polyphonic voices unfolding simultaneously across a shared metrical grid of measures. We map these structural primitives to the EV charging domain as follows:
\begin{itemize}[leftmargin=1.5em]
\item A \textbf{Note} represents an atomic, physically validated charging transaction.
\item A \textbf{Repair Chord} denotes a reconstruction operator that rectifies fragmented data sessions.
\item A \textbf{Harmonic Chord} reflects recurring, localized patterns in residual time-series data.
\item A \textbf{Voice} corresponds to an independent, latent behavioral stream.
\item A \textbf{Measure} is the hierarchical metrical baseline that governs daily, weekly, and holiday cycles.
\item A \textbf{Movement} signifies a macro-level regime shift across the station network.
\end{itemize}

%% ── 1.4 ──────────────────────────────────────────────────────────
\subsection{Contributions and Paper Structure}

Our primary contributions are: (i) \textbf{Axiom-based Falsification Gates} that rigorously validate data properties prior to downstream modeling; (ii) \textbf{$\Gamma$-initialized NMF} with convergence rescaling; (iii) \textbf{Causal De-biasing} via instrument isolation; and (iv) \textbf{Per-Voice Interaction OLS} to mitigate simplex collinearity. 

The remainder of this paper is structured as follows: Section~\ref{sec:architecture} formalizes the Note–Chord–Voice ontology and its axiomatic foundations. Section~\ref{sec:experiment} details the Jiangmen case study, data preprocessing, instrument isolation, and evaluation benchmarks. Section~\ref{sec:results} presents the empirical validation, covering axiom verification, voice separation, movement detection, causal estimates, and revenue simulations. Section~\ref{sec:discussion} confronts theoretical elegance with empirical imperfection, discusses limitations, and outlines generalizability. Section~\ref{sec:conclusion} summarizes the key findings and outlines directions for future work.

%% ======================================================================
\section{The Note–-Chord–-Voice (NCV) Framework Architecture}
\label{sec:architecture}

\subsection{Topological Ontology}
\label{sec:ontology}

Table~\ref{tab:ontology} summarizes the six core concepts of the Note--Chord--Voice framework.

\begin{table}[htbp]
\centering
\caption{Topological definitions for the Note--Chord--Voice framework.}
\label{tab:ontology}
\small
\renewcommand{\arraystretch}{1.4}
\begin{tabularx}{\textwidth}{p{2.2cm} p{1.2cm} >{\RaggedRight\arraybackslash}X p{3.0cm}}
\toprule
\textbf{Concept} & \textbf{Symbol} & \textbf{Definition} & \textbf{Role} \\
\midrule
Note & $n_i$ & $(t_i, E_i, \Delta t_i, p_i, c_i)$ subject to physical manifold $\mathcal{M}$ & Atomic validated event \\
Repair Chord & $C^{r}$ & Information-theoretic reconstruction operator merging fragmented notes when $ \Delta\text{MDL} > 0$ & Data reconstruction \\
Harmonic Chord & $C^{h}$ & Recurrent local tensor motif discovered via Matrix Profile on residuals & Residual pattern discovery \\
Voice & $V_k$ & Horizontal latent stream operationalized as non-negative basis vector in NMF factorization & Source separation \\
Measure & $\Gamma(t)$ & Hierarchical metrical field (e.g. $24\text{ h}$, $168\text{ h}$, and holidays) via STL decomposition & Baseline intensity \\
Movement & $M_j$ & Structural regime change in station load, detected by Foote novelty method & Regime segmentation \\
\bottomrule
\end{tabularx}
\end{table}

\subsubsection{Mathematical Formulations}

\paragraph{Physical Manifold}
An atomic note $n_i = (t_i, E_i, \Delta t_i, p_i, c_i)$ represents a recorded transaction starting at $t_i$, transferring energy $E_i$, over physical duration $\Delta t_i$, with unit energy price $p_i$, and coupon discount $c_i$. A note is considered valid if and only if it belongs to the physical manifold $\mathcal{M}$:
\begin{equation}
\label{eq:manifold_definition}
\mathcal{M} = \left\{ (E_i, \Delta t_i) \in \mathbb{R}_+^2 \;\middle|\; \frac{E_i}{\Delta t_i} \le \eta_{\text{phys}} \cdot P_{\max, i} \right\}
\end{equation}
where $P_{\max, i}$ is the rated peak power of the physical charging pile associated with note $i$, and $\eta_{\text{phys}} = 1.2$ is a tolerance factor that accounts for transient grid fluctuations.

\paragraph{Repair Chord Contraction via Minimum Description Length}
Let $\mathcal{N}$ represent a temporal sequence of notes recorded on a single physical charging pile. A Repair Chord contraction $C^r$ is an information-theoretic operator mapping two adjacent notes $n_i, n_{i+1}$ separated by a time gap $\tau_i = t_{i+1} - (t_i + \Delta t_i)$ to a single merged note $n_{merged} = (t_i, E_i + E_{i+1}, \Delta t_i + \tau_i + \Delta t_{i+1})$. This merging operation is executed if and only if the Minimum Description Length (MDL) change $\Delta \text{MDL}$ is positive:
\begin{equation}
\Delta \text{MDL} = \left[ L(\mathcal{D}_{split} | \mathcal{M}_{split}) + L(\mathcal{M}_{split}) \right] - \left[ L(\mathcal{D}_{merge} | \mathcal{M}_{merge}) + L(\mathcal{M}_{merge}) \right] > 0
\end{equation}
where $L(\mathcal{D} | \mathcal{M})$ represents the negative log-likelihood of the data given the distribution model, and $L(\mathcal{M})$ is the parameter description length. For the MDL computation, both the energy delivered and the physical duration are modeled as Gamma distributions fitted via maximum likelihood estimation on the full session population. The merge decision evaluates the joint log-likelihood under a single-session model (one Gamma pair) versus a two-session model (two independent Gamma pairs), penalized by a chain-length-dependent complexity term $\omega$ that discourages overly aggressive chaining.

\paragraph{Harmonic Chords and the Matrix Profile}
Let $X_s(t)$ be the residual load series of station $s$ after removing the long-term trend and seasonal cycles. We evaluate subsequences over a sliding window of length $m$. The Matrix Profile $P \in \mathbb{R}^{T - m + 1}$ is a vector containing the Euclidean distances between each subsequence and its nearest non-overlapping neighbor in $X_s(t)$, providing an efficient computational framework for time series motif discovery~\cite{yehMatrixProfileAll2016}. A Harmonic Chord $C^h$ is defined as a pair of subsequences $(X[i : i+m], X[j : j+m])$ whose Matrix Profile distance $P[i]$ ranks among the top-$k$ smallest values (with $k=5$ in our implementation), subject to a mutual exclusion zone:
\begin{equation}
P[i] \in \text{top-}k(P) \quad \text{and} \quad |i - j| > 0.25\,m
\end{equation}
where the exclusion zone of $0.25\,m$ prevents trivially overlapping motif pairs while allowing nearby recurrences.

\paragraph{NMF with $\Gamma$-initialization}
Let $\mathbf{X} \in \mathbb{R}_+^{S \times 168}$ be the station-by-hour-of-week energy load matrix. We factorize $\mathbf{X}$ into a station-responsibility matrix $\mathbf{W} \in \mathbb{R}_+^{S \times K}$ and a voice temporal basis matrix $\mathbf{H} \in \mathbb{R}_+^{K \times 168}$ by minimizing:
\begin{equation}
\min_{\mathbf{W}, \mathbf{H} \ge 0} \|\mathbf{X} - \mathbf{W}\mathbf{H}\|_F^2
\end{equation}
via standard multiplicative-update rules. The basis matrix $\mathbf{H}$ is initialized using a $\Gamma$-clustering procedure: per-station 168-hour seasonal profiles $\Gamma_s$ are extracted via STL, then clustered into $K$ groups via $k$-means, and each row $\mathbf{h}_k$ of $\mathbf{H}$ is set to the centroid of cluster $k$. This initializes the optimization within a physically meaningful coordinate space where each voice initially represents a distinct station behavioral archetype. Additionally, the input matrix is min-max normalized to $[0,1]$ before decomposition to improve convergence stability, and subsequently rescaled back to the original scale.

\paragraph{Movement Detection via Foote Novelty}
For each station $s$, we construct the daily load matrix $\mathbf{Y}_s \in \mathbb{R}_+^{D \times 24}$, where $D$ is the total number of operational days. We compute the daily self-similarity matrix $\mathbf{S} \in [0, 1]^{D \times D}$ using cosine similarity. A checkerboard Gaussian kernel matrix $\mathbf{K}$ of size $2L \times 2L$ is convolved along the diagonal of $\mathbf{S}$ to generate the raw novelty curve $N(t)$. Movements $M_j$ are defined as local maxima of the globally normalized novelty curve $N^*(t)$ that exceed a relative prominence threshold, with a minimum inter‑peak distance. This adapts the Foote novelty metric, originally designed for audio segmentation, to detect structural regime shifts in multivariate time series~\cite{footeAutomaticAudioSegmentation2000, leroiRevolutions2020a}. (The specific parameter values used in our pilot are reported in Section~\ref{sec:results}.)

\subsection{Axiomatic Falsification Gates and Graceful Degradation}
\label{sec:axioms}

\textbf{Design Philosophy.} Complex models should not be constructed upon invalid data premises. Before executing downstream routines, the framework implements rigorous falsification tests. If an axiom fails, the pipeline degrades gracefully rather than propagating downstream errors or unfounded distortions. Instead of aborting the analysis, a failed axiom forces a reduction in modeling assumptions and demands a more conservative interpretation of estimates. For instance, if the weekly periodicity axiom (G3) fails, the framework automatically restricts the harmonic basis to daily cycles, preventing spurious 168-hour structures from biasing causal estimates.

Table~\ref{tab:axioms} summarizes each axiom alongside its formal statement, statistical test, pass condition, and empirical outcome on the Jiangmen dataset.

\begin{table}[htbp]
\centering
\caption{Axiom falsification gates: statement, test, pass condition, and result on the Jiangmen dataset.}
\label{tab:axioms}
\small
\setlength{\tabcolsep}{4pt}
\renewcommand{\arraystretch}{1.4}
\begin{tabularx}{\textwidth}{
@{} 
>{\bfseries\RaggedRight\arraybackslash}p{2.2cm}  
>{\raggedright\arraybackslash}X                  
>{\raggedright\arraybackslash}X                  
>{\centering\arraybackslash}p{1.8cm}             
>{\raggedright\arraybackslash}X                  
@{}}
\toprule
Axiom & Statement & Test & Pass Cond. & Result \\
\midrule

A1 Physical Manifold 
& A non-trivial fraction of sessions obey $E/\Delta t \le 1.2 P_{\max}$ 
& Binomial proportion: $H_0$: $>$95\% violate 
& $p < 0.05$ 
& \PASS{} (0.03\% violations, $p \approx 0$) \\

A2 Topological Contractibility 
& Short-gap sessions show MDL compression when merged 
& Wilcoxon on $\Delta\text{MDL}$ for gap $\in (0,2]$ 
& med $> 0$, $p < 0.01$ 
& \PASS{} (median log-likelihood ratio $= +0.43$, $p = 2.3 \times 10^{-25}$) \\

A3 Motif Recurrence 
& Residuals after $\Gamma(t)$ contain recurring patterns 
& Permutation test on MP distance vs.\ phase-randomized surrogate 
& $p < 0.05$ 
& \PASS{} (best $p = 0.000$ over 9 tests) \\

G3 Multi-periodicity 
& Load exhibits both 24\,h and 168\,h periodicity 
& FFT: rank of 24\,h and 168\,h frequency bins $\le 3$ 
& both 
& \FAIL{} (only 24\,h strong; 168\,h rank $\ge 7$) \\

A4 Source Separability 
& NMF outperforms station-mean baseline 
& Bootstrap error + ARI stability 
& err $<$ base, ARI $> 0.7$ 
& \PASS{} ($R^{2} = 0.9921$, stable $K=5$) \\

A5 Interaction Structure 
& (a) Station--hour interaction $>5\%$ (b) NMF stable to treatment inclusion 
& (a) Two-way ANOVA (b) Control--full NMF alignment $r \ge 0.7$ 
& both 
& \PARTIAL{} (16.9\% interaction; 3/5 stable, 2 td) \\

G10 Coupon Relevance 
& Discount rate explains meaningful price variation 
& First-stage $R^{2}$ of discount\_rate $\to$ price\_discount\_pct 
& $R^{2} > 0.05$ 
& \PASS{} ($R^{2} = 0.773$) \\
\bottomrule
\end{tabularx}
\end{table}

\subsubsection{Detailed Axiom Formalism}

\paragraph{A1 -- Physical Manifold Validation}
Let $x_i = \mathbb{I}((E_i, \Delta t_i) \notin \mathcal{M})$ be a binary indicator of a physical violation. We perform a binomial proportion test against a high-contamination null:
\begin{equation}
H_0: p = P(x_i = 1) \ge 0.95 \quad \text{vs.} \quad H_1: p < 0.95
\end{equation}
Failure to reject $H_0$ would imply that at least 95\% of sessions violate the physical manifold, rendering any reconstruction framework inapplicable. In practice, the observed violation rate of 0.03\% yields $p \approx 0$, permitting the decisive rejection of $H_0$.

\paragraph{A2 -- Topological Contractibility}
We assess whether adjacent sessions separated by a short time gap ($\text{gap} \in (0, 2]$ minutes) exhibit physical and informational compression when merged. Let $d_i = \Delta \text{MDL}_i$ be the change in Minimum Description Length for candidate pair $i$. We conduct a one-sided Wilcoxon signed-rank test:
\begin{equation}
H_0: \text{Median}(d_i) \le 0 \quad \text{vs.} \quad H_1: \text{Median}(d_i) > 0
\end{equation}

\paragraph{A3 -- Motif Recurrence}
Let $d_{\min}$ be the minimum value of the Matrix Profile $P$. We generate 200 phase-randomized surrogate time series using Fourier phase randomization, which preserves the power spectrum of the original series while disrupting its temporal structure. Let $\tilde{d}_{\min}^{(r)}$ be the minimum Matrix Profile distance of the $r$-th surrogate. The empirical $p$-value is defined as:
\begin{equation}
p = \frac{1}{200} \sum_{r=1}^{200} \mathbb{I}\left(\tilde{d}_{\min}^{(r)} \le d_{\min}\right)
\end{equation}
We reject $H_0$ (no structural motifs) if $p < 0.05$.

\paragraph{G3 -- Multi-periodicity FFT Gate}
We compute the power spectral density $S(f)$ of the aggregate load series using the Fast Fourier Transform (FFT). Let $f_{24} = 1/24\,\text{h}^{-1}$ and $f_{168} = 1/168\,\text{h}^{-1}$. We sort the discrete frequency bins by spectral energy in descending order. Let $\text{Rank}(f)$ denote the rank position of frequency $f$ in the sorted energy spectrum:
\begin{equation}
\text{Pass Condition: } \max \left( \text{Rank}(f_{24}), \text{Rank}(f_{168}) \right) \le 3
\end{equation}

\paragraph{A4 -- Source Separability}
Let $\mathbf{e}_{nmf}$ be the reconstruction error of NMF with $K$ components, and $\mathbf{e}_{base}$ be the error of a baseline station-mean model. We compute the Adjusted Rand Index (ARI) of the NMF cluster assignments across 50 bootstrap runs. The pass condition requires:
\begin{equation}
\mathbf{e}_{nmf} < \mathbf{e}_{base} \quad \text{and} \quad \text{ARI} > 0.70
\end{equation}

\paragraph{A5 -- Interaction and Stability Gate}
We perform a two-way ANOVA on the station-hour-of-week energy matrix. Let $SS_{\text{interaction}}$ be the sum of squares of the station-by-hour interaction term, and $SS_{\text{total}}$ be the total sum of squares. For stability, let $\mathbf{H}_C$ be the basis matrix extracted from the control data (no treatment), and $\mathbf{H}_F$ be the basis matrix from the full dataset. We calculate the maximum cross-correlation $r_k$ for each voice $k$ via optimal assignment:
\begin{equation}
r_k = \max_j \text{corr}(\mathbf{h}_{C,k}, \mathbf{h}_{F,j})
\end{equation}
A voice is classified as \emph{stable} if $r_k \ge 0.70$, as \emph{treatment-driven} if $r_k < 0$, and as \emph{unstable} if $0 \le r_k < 0.70$. Only stable voices support causal identification; treatment-driven voices remain strictly descriptive; unstable voices signal inadequate source separation. In the Jiangmen dataset, no voices fell into the unstable category. The overall pass condition requires a majority of voices to be stable:
\begin{equation}
n_{\text{stable}} \ge \left\lfloor \frac{K}{2} \right\rfloor + 1
\end{equation}
If this condition is met while containing treatment-driven voices, the outcome is qualified as \emph{partial}, restricting causal inference exclusively to stable components.

\paragraph{G10 -- Coupon Relevance Gate}
We estimate the first-stage instrument equation: $\text{price\_discount\_pct}_i = \alpha + \theta \cdot \text{discount\_rate}_i + \nu_i$. The pass condition requires the first-stage $R^2$ to exceed a minimum threshold:
\begin{equation}
R^2 > 0.05
\end{equation}
In practice, the estimated $R^2 = 0.773$ far exceeds this threshold, confirming that the discount rate accounts for substantial price variance.

\subsection{Pipeline Execution Strategy and Key Design Decisions}
\label{sec:pipeline}

The pipeline consists of six sequential steps (Steps 0--5), complemented by an intermediate axiom gate (Step~0.5), as illustrated in Figure~\ref{fig:pipeline}. The following design decisions are embedded directly into the execution flow:

\begin{figure}[htbp]
\centering
\begin{tikzpicture}[
node distance=1.1cm,
box/.style={draw, rounded corners, fill=blue!8, minimum width=2.8cm, minimum height=0.8cm, font=\small, align=center},
gate/.style={draw, diamond, fill=orange!15, aspect=2.2, font=\scriptsize, inner sep=2pt, align=center},
arr/.style={->, thick, >=stealth}
]
\node[box] (s0) {Step 0: Preprocess};
\node[gate, below=of s0] (ax) {Step 0.5\\Axiom Gate};
\node[box, below left=1.2cm and 0.4cm of ax] (s1) {Step 1: Repair Chords};
\node[box, below right=1.2cm and 0.4cm of ax] (s2) {Step 2: Harmonic Chords};
\node[box, below=2.2cm of ax] (s3) {Step 3: Voice Separation};
\node[box, below=of s3] (s4) {Step 4: Movements};
\node[box, below=of s4] (s5) {Step 5: Causal \& Pricing};

\draw[arr] (s0) -- (ax);
\draw[arr] (ax) -| node[pos=0.25, above, font=\scriptsize]{A1, A2} (s1);
\draw[arr] (ax) -| node[pos=0.25, above, font=\scriptsize]{G3, A3} (s2);

% Fixed: changed to pos=0.5, below to place text safely in the open gap
\draw[arr] (s1) |- node[pos=0.5, below, font=\scriptsize]{Repaired Notes} (s3);
\draw[arr] (s2) |- node[pos=0.5, below, font=\scriptsize]{$\Gamma$ Profiles} (s3);

\draw[arr] (s3) -- node[right, font=\scriptsize]{NMF Factors} (s4);
\draw[arr] (s4) -- node[right, font=\scriptsize]{Changepoints} (s5);
\end{tikzpicture}
\caption{Note--Chord--Voice pipeline overview. The Axiom Gate (Step~0.5) tests data suitability before any complex model is built, ensuring graceful degradation if assumptions fail.}
\label{fig:pipeline}
\end{figure}
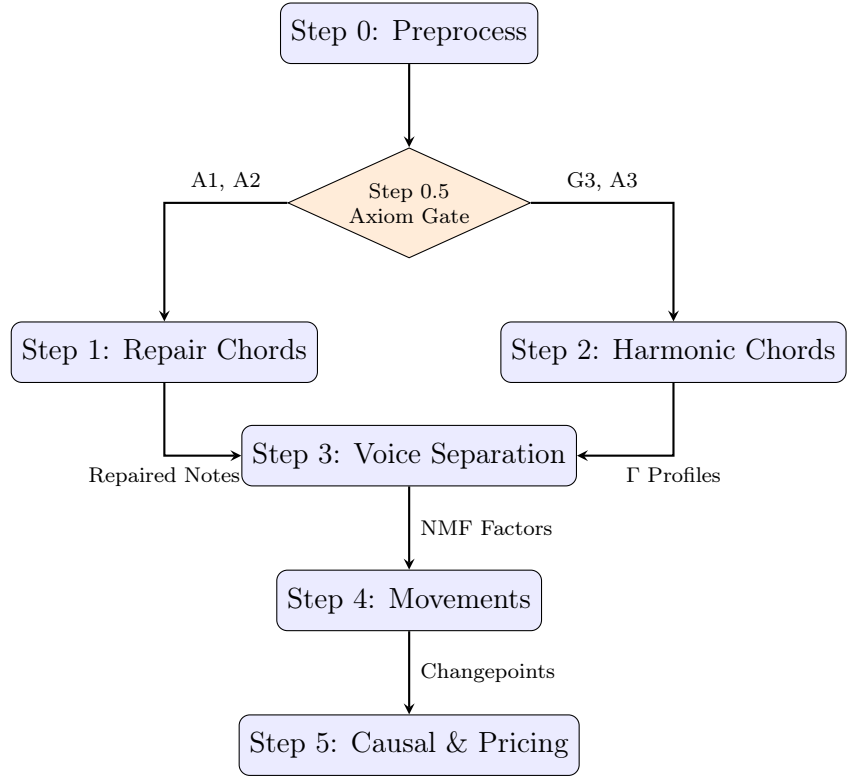

\begin{enumerate}[leftmargin=1.5em]
\item \textbf{Step 0: Preprocessing.} Remove physically impossible sessions ($\text{start} \ge \text{end}$, $E \le 0$), encode holidays, parse promotion names into structured tags, and audit the physical manifold. \emph{Output: 495,707 cleaned sessions.}
\item \textbf{Step 0.5: Axiom Gate.} Execute falsification tests A1--A5, G3, and G10, select $K=5$ latent voices, and classify voices as stable $\{0, 2, 3\}$ or treatment-driven $\{1, 4\}$.
\item \textbf{Step 1: MDL Repair Chord Contraction.} Merge fragmented notes using a greedy stack algorithm subject to physical manifolds, a log-likelihood ratio (LR) pre-filter, and MDL gain gates. MDL is bounded to gaps $\le 2$ minutes because positive log-likelihood evidence concentrates exclusively in the $(0,2]$ minute bracket; longer gaps likely reflect distinct individual sessions (see Section~\ref{sec:results}). \emph{Result: 495,707 sessions reduced to 490,362 notes via 5,345 successful merges.}
\item \textbf{Step 2: STL Decomposition \& Harmonic Motif Discovery.} Extract per-station seasonal baselines $\Gamma(t)$ via seasonal and trend decomposition using LOESS (STL) and compute the Matrix Profile on residual series. \emph{Result: 210 motifs successfully pass the MDL gate.}
\item \textbf{Step 3: Voice Separation \& Duration Model.} Perform NMF on the station-by-hour-of-week energy matrix using $\Gamma$-initialization. $\Gamma$-initialized NMF leverages STL measures to establish a physically meaningful coordinate origin for $\mathbf{H}$, while input normalization stabilizes multiplicative updates. The duration model incorporates a constant current-constant voltage (CC-CV) lower bound (assuming charging efficiency $\eta_{\text{charge}} = 0.90$), a softplus overhead term, and an idle component. \emph{Result: NMF $R^{2} = 0.9921$; duration model $R^{2} = 0.5409$.}
\item \textbf{Step 4: Movement Inference \& Load Forecasting.} Apply the Foote novelty method using $L=14$ days, a $30\%$ prominence threshold, a minimum inter-peak distance of $28$ days, and global normalization. \textbf{Decision rationale:} Global normalization ensures that novelty scores remain comparable across stations with varying load magnitudes. \emph{Result: 96 changepoints detected (15 major structural shifts); WMAPE improved by 15.5\%.}
\item \textbf{Step 5: Causal \& Pricing Layer.} Execute coupon grading (A/B/C/D/None), estimate first-stage effects, fit per-voice OLS models, and run revenue simulations. \textbf{Decision rationale:} Fitting separate OLS regressions per voice avoids severe collinearity induced by the simplex constraint $\sum_k r_{ik}=1$. Furthermore, the voice-weighted sensitivity threshold (combined responsibility $>0.5$) establishes a hard decision boundary for revenue simulations while retaining continuous soft memberships.
\end{enumerate}

\subsection{Causal Decoupling Formulation}
\label{sec:causal_decoupling}

\subsubsection{The Collider Bias Problem}
Conventional approaches to EV price elasticity typically group charging sessions based on total energy delivery or connection duration, making cluster-wise elasticity estimation unreliable. Let $T$ be the treatment (coupon discount), $Y$ be the outcome (connection duration), $U$ be unobserved user preference (e.g., price sensitivity, urgency), and $C$ be the cluster assignment.

Because $C$ is constructed directly from post-treatment outcomes, it acts as a collider. Conditioning on $C$ opens a backdoor path $T \rightarrow Y \leftarrow U$, inducing endogenous selection bias. The Note--Chord--Voice framework bypasses this structural artifact by defining soft, pre-treatment latent behavioral dimensions (Voices) based on station-level aggregate load profiles $\mathbf{W}$, rather than session-level post-treatment outcomes.

\subsubsection{Per-Voice Interaction Model}
Let $r_{ik}$ denote the NMF responsibility (membership) of session $i$ in voice $k$, where the responsibilities reside on the probability simplex:
\begin{equation}
\sum_{k=1}^K r_{ik} = 1 \quad \forall i
\end{equation}
When estimating a pooled interaction regression:
\begin{equation}
y_i = \alpha + \sum_{k=1}^K \beta_k \left( r_{ik} \cdot T_i \right) + \sum_{k=1}^K \gamma_k r_{ik} + \epsilon_i
\end{equation}
The simplex constraint induces perfect multicollinearity with the intercept term ($\sum_k r_{ik} = \mathbf{1}$). To resolve this issue, we fit an independent OLS regression for each voice $k$ on the full sample of treated and control sessions:
\begin{equation}
\label{eq:per_voice_ols}
y_i = \alpha_k + \beta_{1k}\, r_{ik} + \beta_{2k}\, T_i + \beta_{3k}\, (r_{ik} \times T_i) + \beta_{4k}\, E_i + \epsilon_i
\end{equation}
The coefficient $\beta_{3k}$ on the interaction term isolates the voice-specific heterogeneous treatment effect: how the price response varies with the degree of membership in voice $k$. This approach avoids simplex collinearity while controlling for the main effects of voice membership, treatment status, and energy demand.

\noindent\textbf{Theoretical requirement.} Robust causal identification of $\beta_{3k}$ requires a quasi-random instrument that is orthogonal to unobserved confounders (such as time-of-day preferences or user urgency). The construction of such an instrument from promotional coupon tags is detailed in Section~\ref{sec:instrument}.

%% ======================================================================
\section{Experimental Setup and Case Study Design}
\label{sec:experiment}

\subsection{Dataset Description}
\label{sec:dataset}

We validate the NCV framework on a real-world dataset from Jiangmen, Guangdong Province, southern China. The dataset spans 20 public EV charging stations over the period July 2024 -- March 2025 (9 months), containing 495,707 recorded charging sessions after initial cleaning. Each session records the start time, delivered energy (kWh), physical connection duration (minutes), unit price (CNY/kWh), and associated promotional discount. The physical charger infrastructure includes a mix of AC slow chargers (7\,kW), DC fast chargers (30--120\,kW), and a small number of ultra-fast chargers (up to 480\,kW). Rated peak power $P_{\max}$ for each charger is documented, enabling the physical manifold constraint defined in Eq.~\eqref{eq:manifold_definition}.

\subsection{Preprocessing and Instrument Isolation: The A/B/C/D Coupon Taxonomy}
\label{sec:instrument}

A critical step for robust causal identification is isolating price variation that is orthogonal to underlying charging behavior. We parsed raw promotion names into four distinct categories, as summarized in Table~\ref{tab:coupon_grades}.

\begin{table}[htbp]
\centering
\caption{Coupon grading taxonomy for Jiangmen promotions.}
\label{tab:coupon_grades}
\small
\renewcommand{\arraystretch}{1.3}
\begin{tabularx}{\textwidth}{
l 
>{\RaggedRight\hsize=1.3\hsize}X 
>{\RaggedRight\hsize=1.1\hsize}X 
>{\RaggedRight\hsize=0.6\hsize}X
}
\toprule
\textbf{Grade} & \textbf{Description} & \textbf{Identification rationale} & \textbf{Role} \\
\midrule
Grade A & Generic platform-wide coupons, no targeting, available to all users & Exogenous: orthogonal to session timing and user characteristics & Primary instrument \\
\addlinespace
Grade B & Night-time specific coupons (e.g., ``off-peak discount'') & Endogenous: severely confounded by circadian timing & Excluded from main analysis \\
\addlinespace
Grade C & Targeted promotions (user-specific, post-purchase, station-specific) & Correlated with user behavior and station characteristics & Excluded \\
\addlinespace
Grade D & Other/unknown promotional tags & Ambiguous; excluded to maintain instrument purity & Excluded \\
\addlinespace
None & No discount applied & Natural control group & Control \\
\bottomrule
\end{tabularx}
\end{table}

Our primary causal analysis compares Grade A treated sessions against unpromoted control sessions. Grade A coupons satisfy the quasi-random instrument criteria of Section~\ref{sec:causal_decoupling}: they are issued platform-wide without conditioning on time, location, or user history. A covariate balance analysis (presented later in Section~\ref{sec:causal_results}) confirms that the night-session share is nearly identical between Grade A (18.1\%) and None (17.4\%), while price differs significantly (0.962 vs.\ 1.214 CNY/kWh), reflecting the genuine coupon treatment rather than a temporal confounder. Grade B is excluded because its 86\% night-time concentration would otherwise induce circadian selection bias into the estimated price elasticity.

\subsection{Baseline Models}
\label{sec:baselines}

To establish proof-of-concept validity and isolate the performance gains attributable to the NCV framework, we benchmark its core components against minimalist reference models. The utilization of minimalist baselines is a deliberate methodological choice designed to isolate the structural contributions of each framework component, rather than an absence of state-of-the-art (SOTA) architectures.

\paragraph{Load forecasting baseline}
We employ a \textbf{historical station-mean baseline} that predicts hourly station load using the historical mean load for that specific hour-of-week over the training period.

\paragraph{Duration modeling baseline}
We implement an \textbf{unconstrained time-only linear model} that regresses $\log(\text{duration})$ solely on hour-of-week indicator variables, capturing baseline circadian trends without physical constraints or transaction granularity to establish a performance floor.

\subsection{Evaluation Metrics}
\label{sec:metrics}

Performance is evaluated across four complementary dimensions:

\begin{itemize}[leftmargin=1.5em]
\item \textbf{Load forecasting:} The Weighted Mean Absolute Percentage Error (WMAPE), defined as $\text{WMAPE} = \sum_{s,t} w_{s,t} |\hat{y}_{s,t} - y_{s,t}| / \sum_{s,t} w_{s,t} y_{s,t}$, with weights $w_{s,t}$ proportional to the energy volume of each station.
\item \textbf{Duration modeling:} The coefficient of determination ($R^{2}$), evaluated at both the session level and the station-hour-of-week aggregate level.
\item \textbf{Clustering stability:} The Adjusted Rand Index (ARI) measuring the stability of NMF-based voice assignments across 50 bootstrap resamples.
\item \textbf{Causal effects:} The heterogeneous treatment effect coefficients $\beta_{3k}$ from the per-voice OLS specification, evaluated with heteroskedasticity-robust standard errors at the 5\% significance level.
\end{itemize}

%% ======================================================================
\section{Empirical Validation and Results}
\label{sec:results}

\subsection{Axiom Verification and Manifold Validation}
\label{sec:axiom_results}

\begin{figure}[htbp]
\centering
\includegraphics[width=0.85\textwidth]{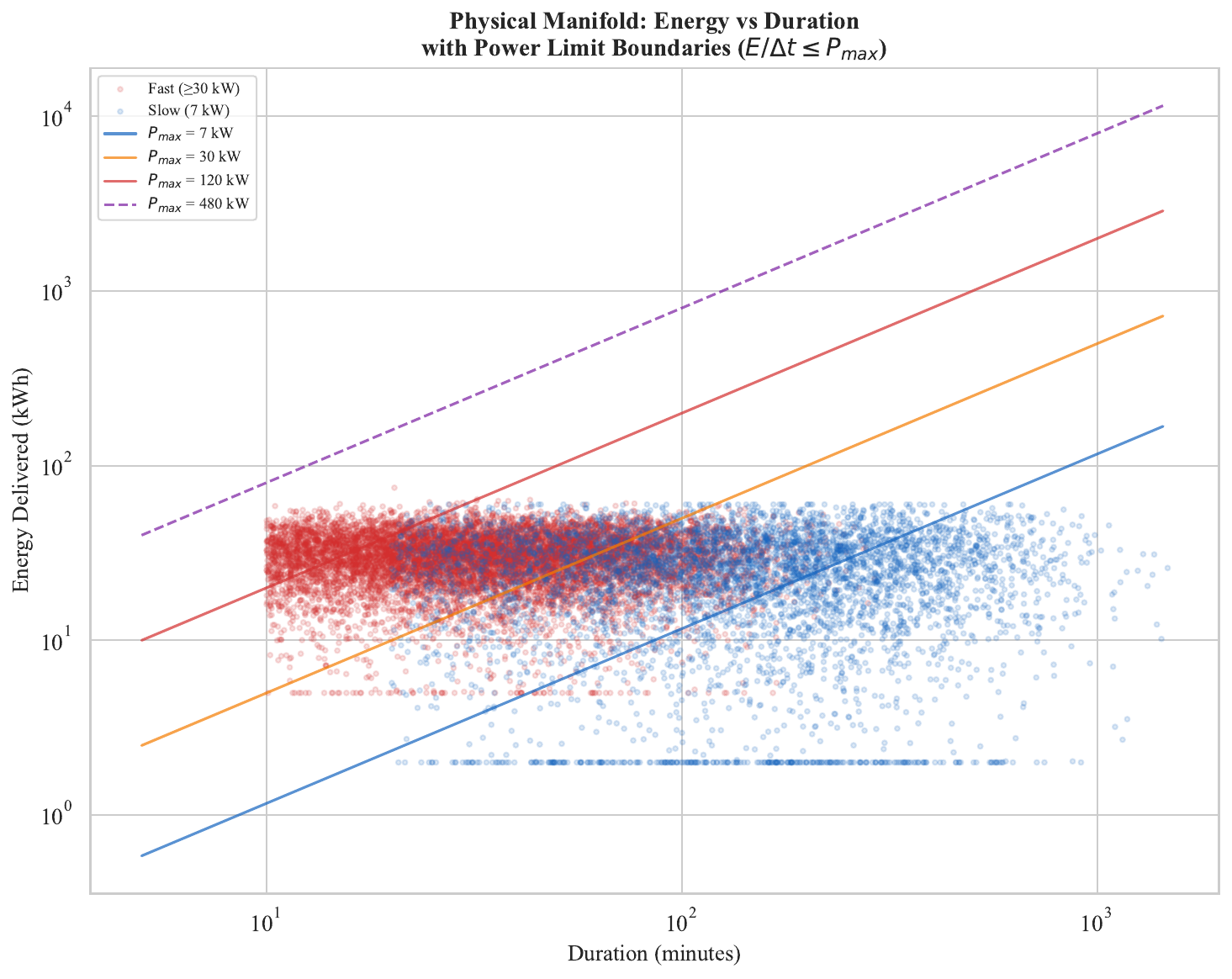}
\caption{Physical manifold: energy vs.\ duration on log--log axes, colored by charger type. Diagonal lines show power limits ($E/\Delta t = P_{\max}$) for 7\,kW, 30\,kW, 120\,kW, and 480\,kW chargers. The $\Gamma$-initialized NMF is fitted to data within these boundaries, while unconstrained models might produce impossible states.}
\label{fig:physical_manifold}
\end{figure}

Table~\ref{tab:axioms} (presented in Section~\ref{sec:axioms}) summarizes the pass/fail status of all falsification gates. The physical manifold validation reveals that 99.97\% of sessions obey $E/\Delta t \le 1.2 P_{\max}$, with only 135 violations (0.03\%). Figure~\ref{fig:physical_manifold} visualizes the energy--duration relationship against power limit boundaries, confirming that virtually all data lie within physically plausible regions.

\begin{figure}[htbp]
\centering
\includegraphics[width=\textwidth]{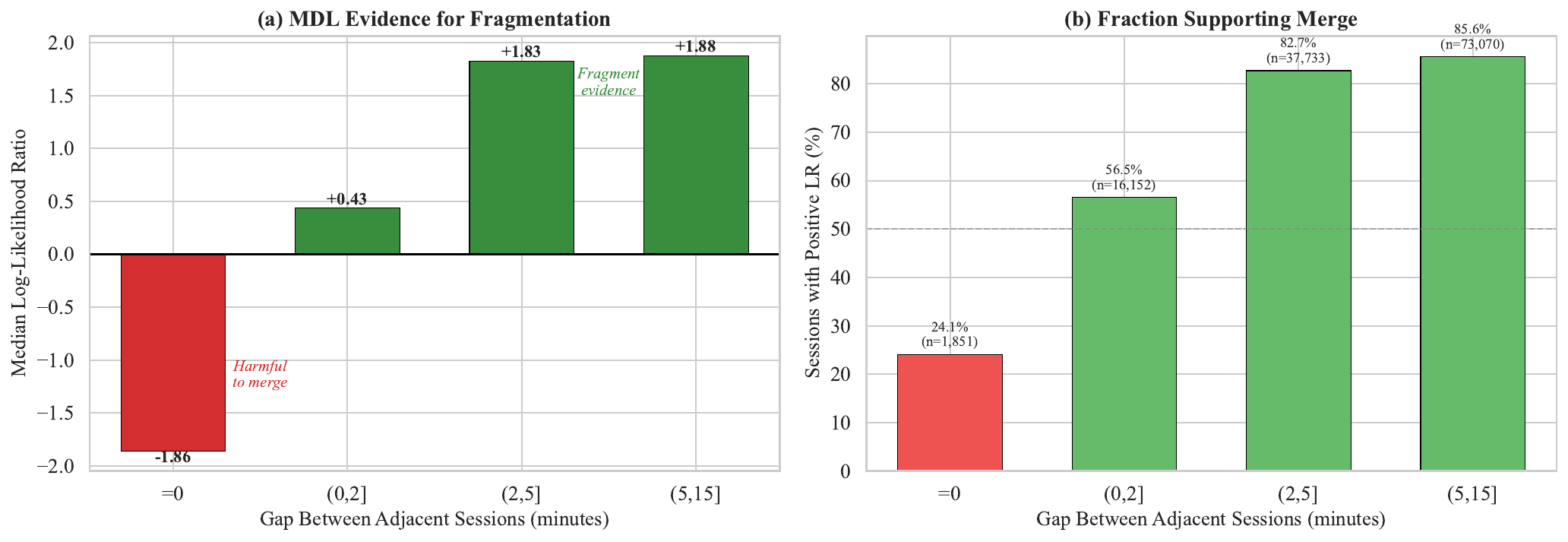}
\caption{A2 axiom test results by gap bracket. Left: Median log-likelihood ratio (negative for back-to-back sessions, positive for short gaps). Right: Fraction of session pairs where merging is statistically beneficial. The fragmentation signal concentrates in the $(0,2]$ minute gap bracket, contradicting the assumption that back-to-back sessions are the primary fragmentation source.}
\label{fig:a2_brackets}
\end{figure}

\textbf{Insight from Axiom A2: Back-to-back sessions lack fragmentation characteristics.} A surprising finding, shown in Figure~\ref{fig:a2_brackets}, is that sessions with zero gap ($\text{gap}=0$) have a negative median log-likelihood ratio (LR $= -1.86$), implying that their consolidation is statistically counterproductive. Positive log-likelihood ratio evidence concentrates strictly within the $(0, 2]$ minute gap bracket (median LR $= +0.43$, $p = 2.3 \times 10^{-25}$). This indicates that the primary fragmentation mechanism stems not from hardware socket timeouts (which would generate a zero-gap distribution), but rather from billing cycle resets or brief user-initiated reconnections. Accordingly, the Repair Chord operator is restricted to gaps of two minutes or less.

\textbf{Graceful degradation under Axiom Gate G3 failure.} While the Fast Fourier Transform (FFT) spectral analysis successfully isolates the 24-hour diurnal cycle, the 168-hour weekly cycle fails to achieve a high spectral rank ($\ge 7$). This indicates that public charging stations in Jiangmen exhibit weak differentiation between weekend and weekday demand profiles. Consequently, the framework executes graceful degradation: while the 168-hour basis is retained for NMF as a structural prior, the $\Gamma$ measure used for harmonic chord extraction restricts its periodic components exclusively to daily cycles. This prevents spurious weekly structure from biasing downstream motif discovery and causal estimation.

\subsection{Voice Separation and Load Modeling}
\label{sec:voice_results}

NMF achieves $R^{2} = 0.9921$ compared to a station-mean baseline of $R^{2} = 0.8475$ ($+0.1446$). To verify robustness against aggregation bias, we re-fitted the NMF model on a row-normalized pivot matrix; the resulting temporal basis vectors ($\mathbf{H}$) and downstream causal metrics remained virtually identical, confirming that the discovered voices represent volume-invariant behavioral chronotypes rather than artifacts of station scale.

\begin{figure}[htbp]
\centering
\includegraphics[width=0.9\textwidth]{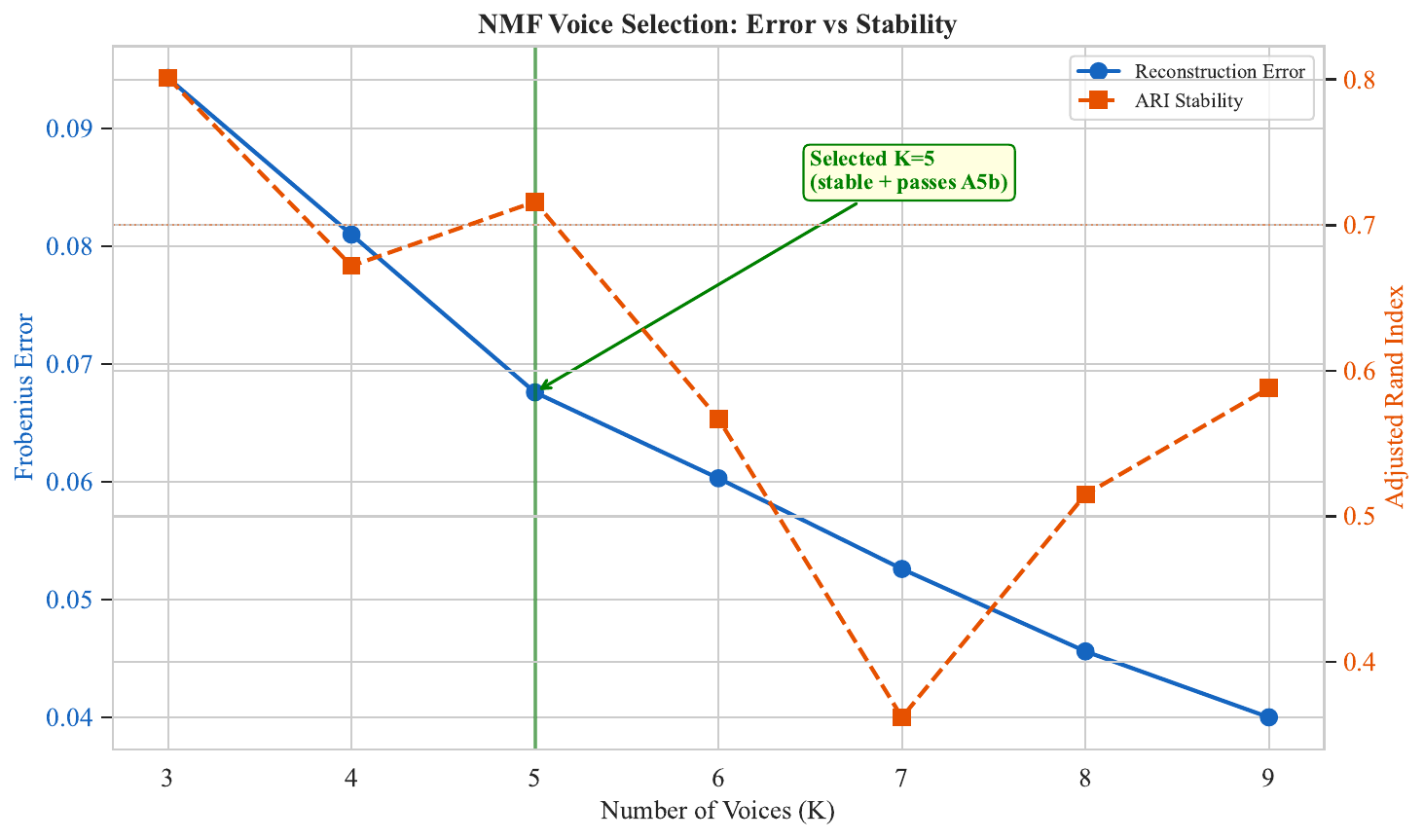}
\caption{NMF voice selection: reconstruction error (left axis) and stability (ARI, right axis) vs.\ number of voices $K$. $K=5$ is selected as the smallest $K$ passing both A4 and A5(b). $K=3$ achieves higher ARI but fails A5(b) (only 1/3 voices stable).}
\label{fig:k_selection}
\end{figure}

Figure~\ref{fig:k_selection} depicts reconstruction error and stability (ARI) as a function of the number of voices $K$. $K=5$ was selected as the smallest $K$ passing both A4 (error $<$ baseline) and A5(b) (stability to treatment inclusion).

\begin{figure}[htbp]
\centering
\includegraphics[width=\textwidth]{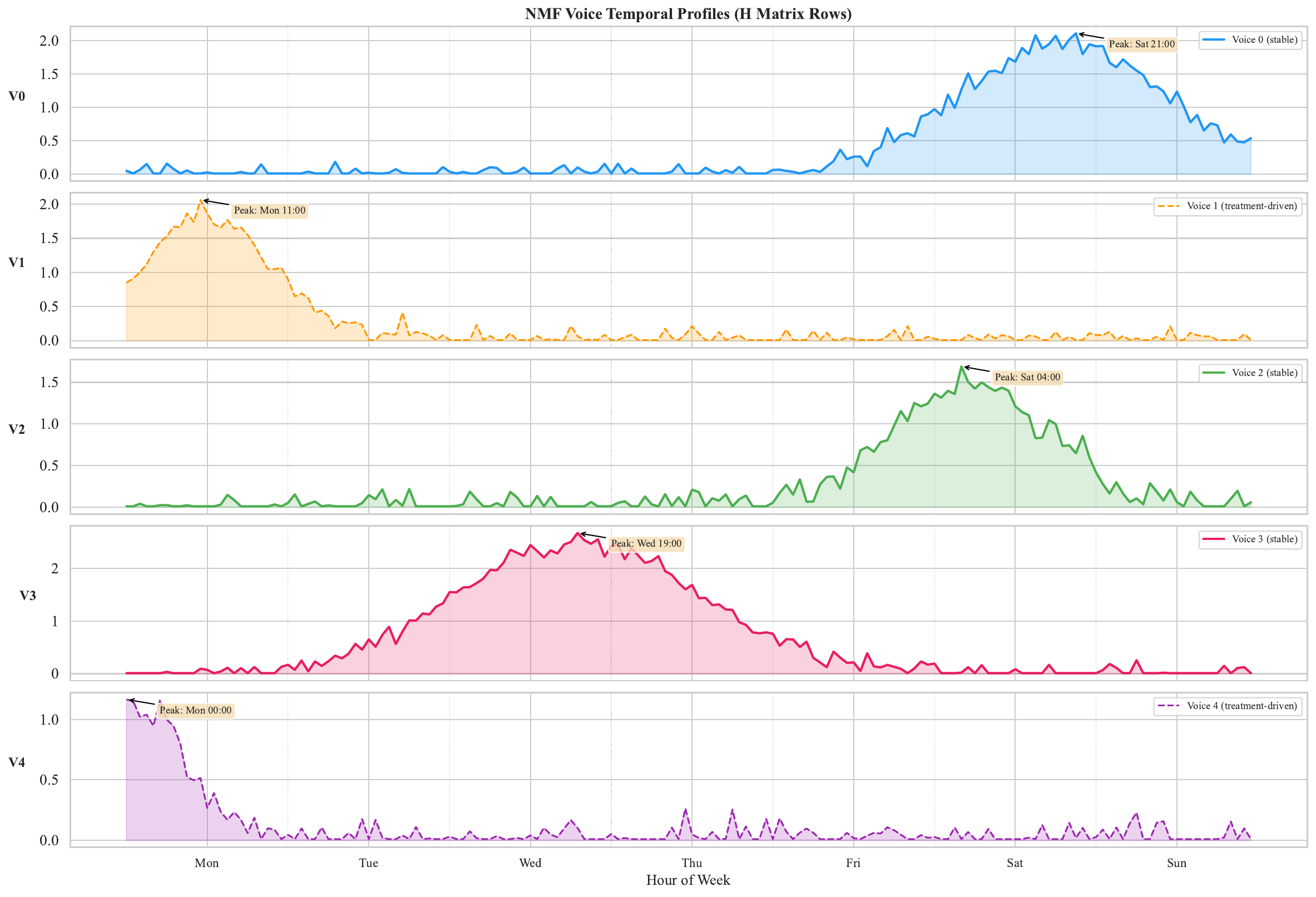}
\caption{NMF voice temporal profiles (rows of $\mathbf{H}$). Each panel shows one voice's normalized energy intensity across 168 hours of the week. Solid lines = stable voices (support causal claims); dashed lines = treatment-driven voices (descriptive only).}
\label{fig:voice_profiles}
\end{figure}

\begin{figure}[htbp]
\centering
\includegraphics[width=\textwidth]{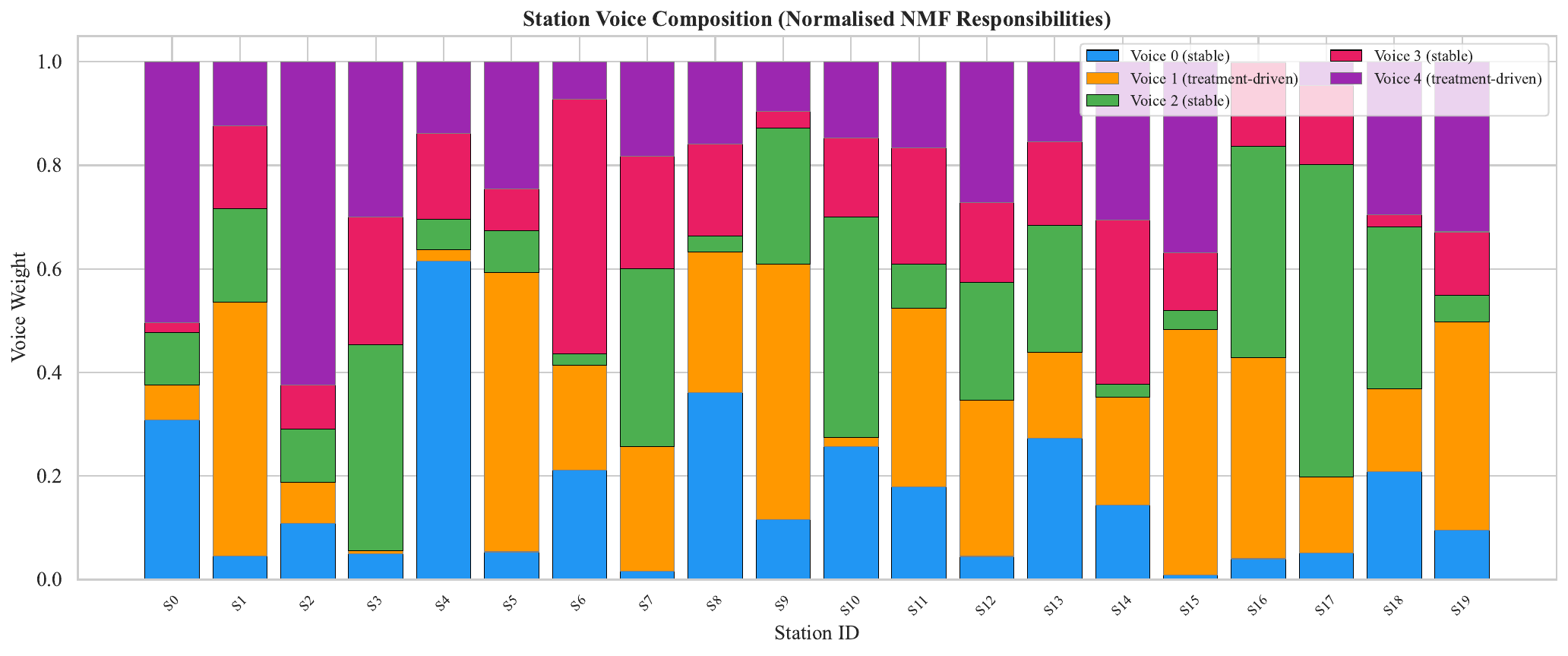}
\caption{Station voice composition (normalized NMF responsibilities). Each bar shows how a station's load is distributed across the 5 voices. Stations dominated by a single voice have clear behavioural signatures; mixed stations reflect overlapping user populations.}
\label{fig:station_voices}
\end{figure}

Figure~\ref{fig:voice_profiles} displays the temporal profiles of the five discovered voices, and Figure~\ref{fig:station_voices} shows the voice composition across stations. Voices~0 and~2 peak on weekends; Voice~3 peaks mid-week evenings. Solid lines denote stable voices (supportive of causal claims); dashed lines denote treatment-driven voices (descriptive only).

\paragraph{Duration model performance.} The time-only baseline achieves $R^{2} = 0.0334$ at the session level. The physically-constrained energy-dependent model---which includes a CC-CV lower bound, a softplus overhead term, and a $\pi$-gated idle component---achieves $R^{2} = 0.5409$ at the station-hour-of-week aggregate level. Although these two $R^{2}$ values are not directly comparable due to differing granularities, the aggregate-level fit is substantially higher because session-level variance is smoothed out. The estimated mean idle probability $\bar{\pi} = 0.597$ indicates that approximately 60\% of voice-hour cells contain a meaningful post-charge idle component.

\subsection{Movement Detection and Forecasting Performance}
\label{sec:movement_results}

\begin{figure}[htbp]
\centering
\includegraphics[width=\textwidth]{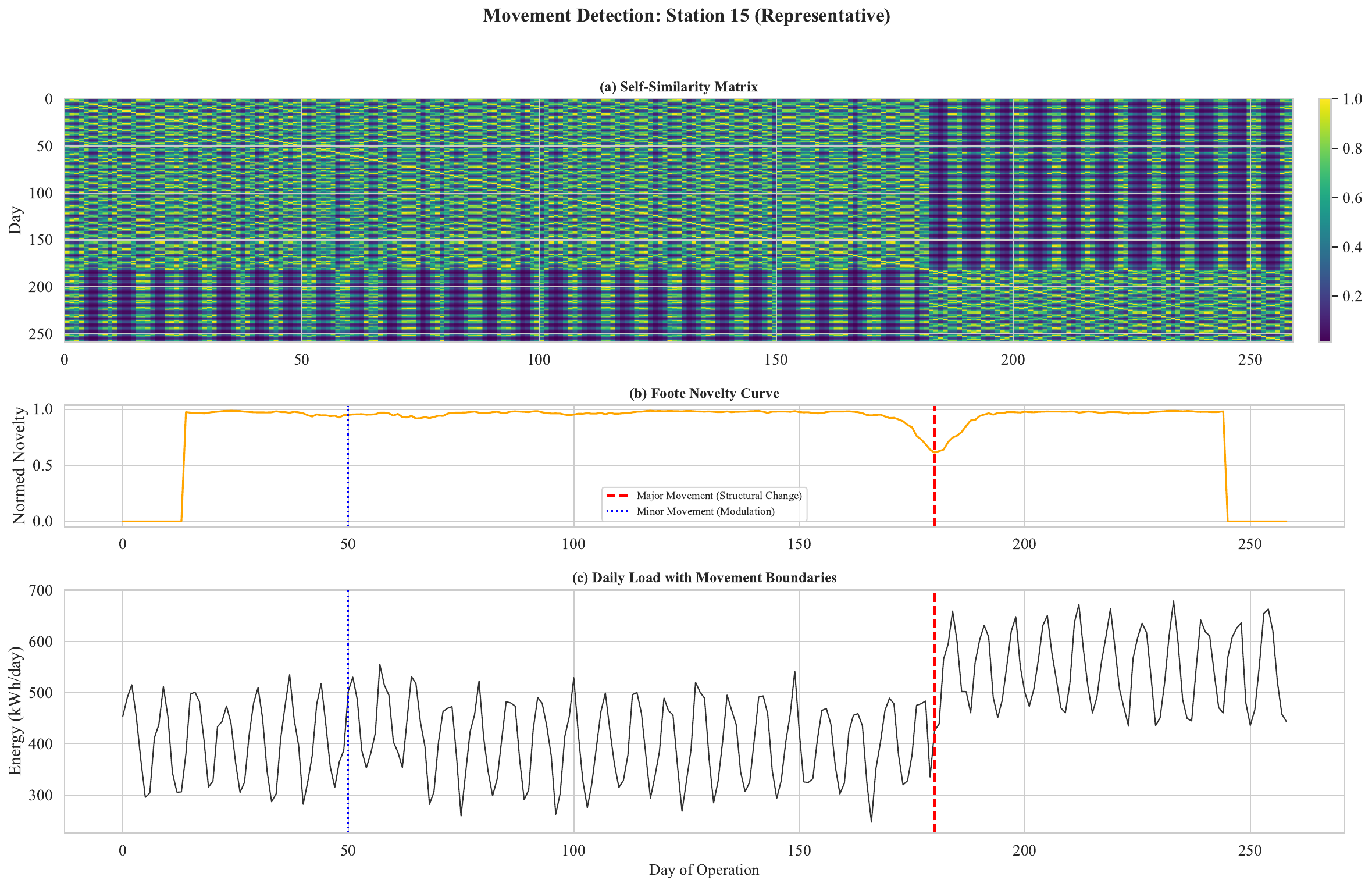}
\caption{Movement detection for a representative station. Top: cosine self-similarity matrix of daily load profiles. Middle: Foote novelty curve (globally normalized) with detected changepoints. Bottom: daily energy load with Major (solid lines) and Minor (dashed lines) Movement boundaries.}
\label{fig:movements}
\end{figure}

Figure~\ref{fig:movements} illustrates movement detection for a representative station. Across all 20 stations, 96 changepoints are detected, of which 15 are classified as major (2 births, 9 structural changes, 4 structural declines). The strongest structural change occurred at Station~2 on September 24, 2024, rising from $390$ to $544\text{ kWh/day}$ ($+40\%$, raw novelty $= 10.76$).

Hourly load forecasting via NMF voice reconstruction achieves $\text{WMAPE} = 0.3257$, compared to the station-mean baseline $\text{WMAPE} = 0.3855$, an improvement of 15.5\%. Although our forecasting evaluation focuses on these structural baselines, the result demonstrates that the voice decomposition captures meaningful temporal structure beyond simple historical averages.

\subsection{Causal Insights and Heterogeneous Treatment Effects}
\label{sec:causal_results}

\begin{figure}[htbp]
\centering
\includegraphics[width=0.9\textwidth]{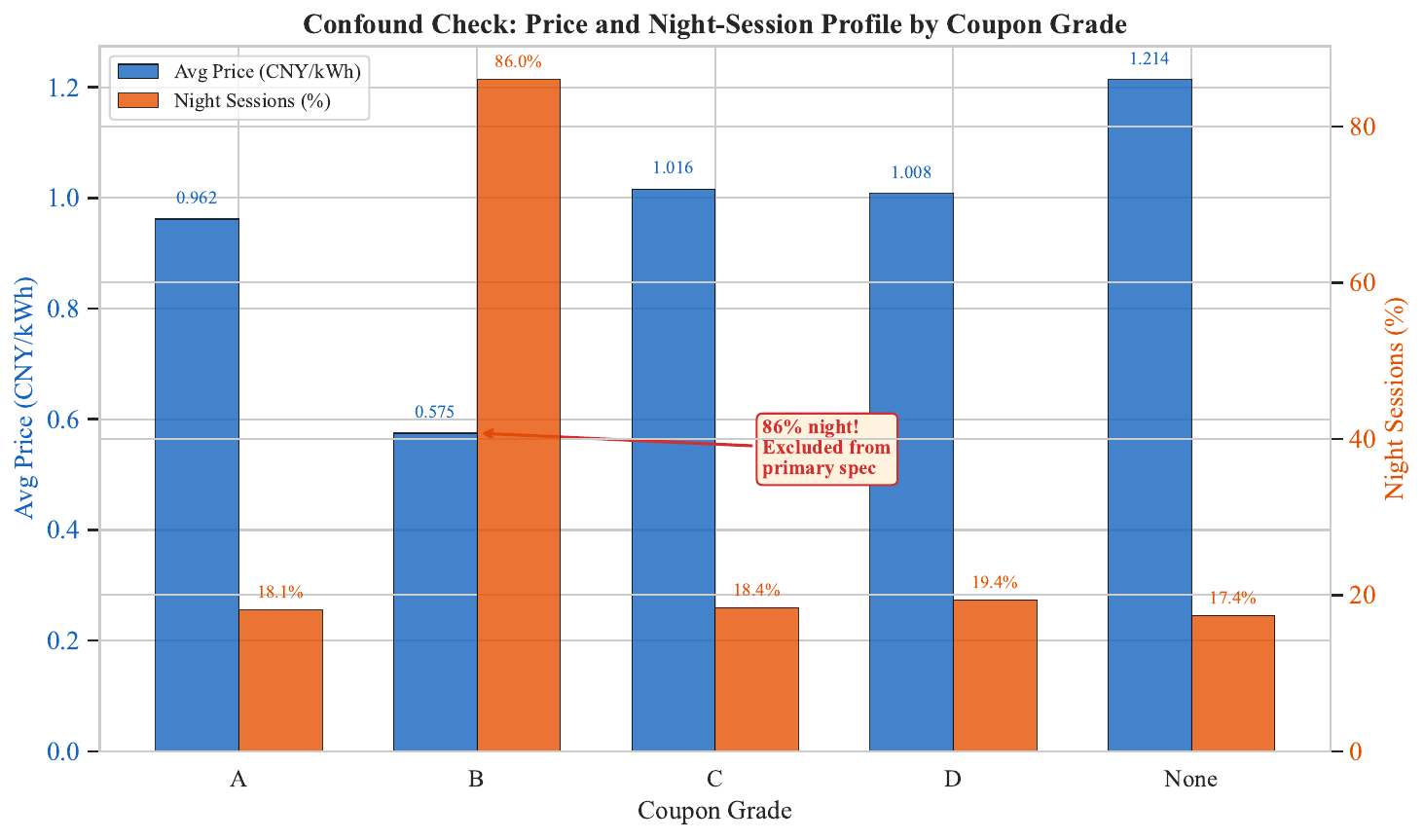}
\caption{Confounder check by coupon grade. Left axis: average price per kWh (blue); right axis: percentage of night-time sessions (orange). Grade~B has an 86\% night-time session concentration and average price 0.575 CNY/kWh, confirming that night-time electricity pricing---not coupons---drives its low price.}
\label{fig:confound_check}
\end{figure}

\paragraph{Confound check.} Figure~\ref{fig:confound_check} visualizes the price and night-time session profile across coupon grades. Grade A closely matches the control group (None) on the pre-treatment temporal confounder (night-session share: 18.1\% vs.\ 17.4\%), confirming its suitability as a quasi-random instrument. The price difference (0.962 vs.\ 1.214 CNY/kWh) reflects the coupon treatment itself rather than an underlying confounder. The 86\% night-time concentration of Grade B sessions validates its exclusion.

\paragraph{First-stage and HTE estimation.} The first-stage regression of price per kWh on treatment status (Grade A vs.\ None), controlling for station and hour-of-day fixed effects, yields a coefficient of $-0.2332$ CNY/kWh ($R^{2} = 0.4483$, $p \approx 0$), satisfying the G10 relevance criterion.

\begin{figure}[htbp]
\centering
\includegraphics[width=0.9\textwidth]{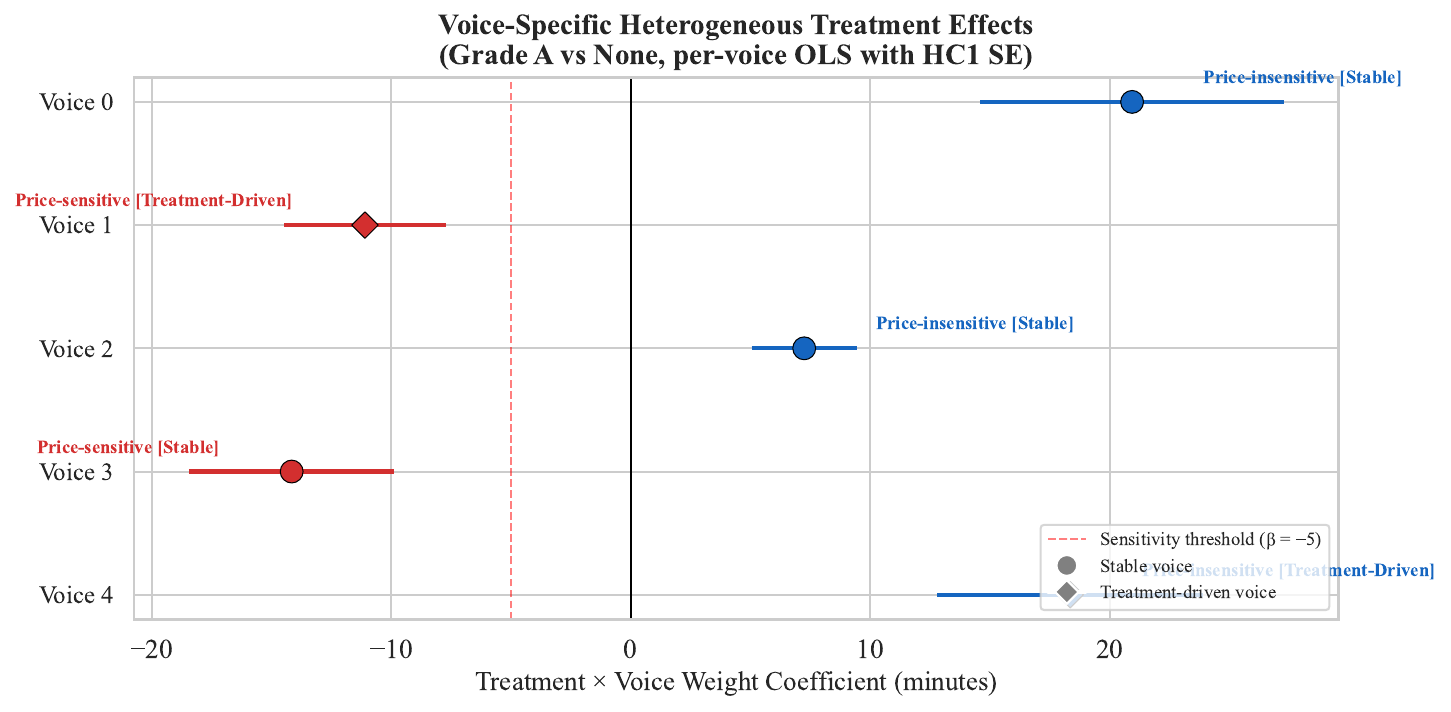}
\caption{Voice-specific heterogeneous treatment effects (Grade~A vs.\ None). Circles = stable voices (support causal claims); diamonds = treatment-driven voices (descriptive only). Red = price-sensitive ($\beta < -5$ minutes, $p < 0.1$); blue = price-insensitive. Voice~3 (stable, $\beta = -14.16$ minutes) is the clean causal finding.}
\label{fig:hte_forest}
\end{figure}

Figure~\ref{fig:hte_forest} presents the voice-specific heterogeneous treatment effects from the per-voice interaction OLS (Eq.~\ref{eq:per_voice_ols}).

\textbf{Key finding.} Voice~3 yields the primary actionable causal result: it is stable (its NMF profile does not shift with treatment inclusion) and price-sensitive ($\beta = -14.16$ minutes, $p < 0.001$). Voice~1 is also price-sensitive but treatment-driven, supporting only descriptive claims. All other voices (0, 2, 4) are not significantly price-sensitive.

\paragraph{Movement-conditional effects (preliminary).} We interacted the voice-specific treatment term with a binary indicator for sessions within $\pm$14 days of a major movement boundary. For Voices 0, 1, 2, and 4, the interaction coefficients were large and statistically significant ($\beta \approx -35$ to $-54$ minutes, $p < 0.001$), while Voice~3 showed no significant interaction ($\beta = -1.60$, $p = 0.53$). However, because movements are detected endogenously from load data that is itself influenced by discounts, these significant interactions likely reflect confounding rather than a causal shift in price sensitivity. We treat these results as preliminary and do not incorporate them into the primary causal interpretation.

\subsection{Revenue Counterfactuals and Policy Implications}
\label{sec:revenue}

\begin{figure}[htbp]
\centering
\includegraphics[width=0.9\textwidth]{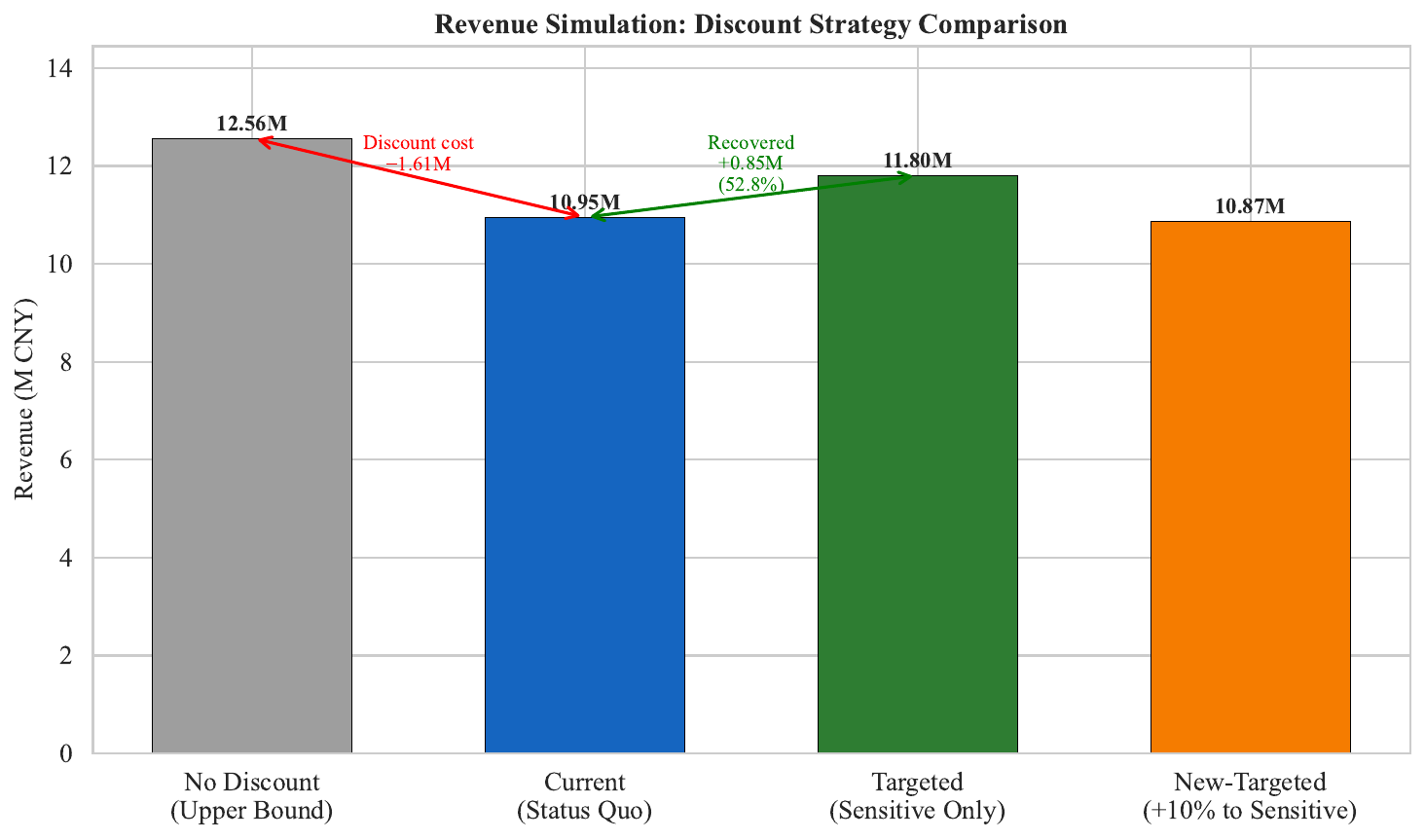}
\caption{Revenue counterfactuals under alternative discount strategies. Comparison of baseline expenditure, targeted allocation to price-sensitive voices (Voices 1 and 3), and the new-targeted expansion scenario, illustrating net expenditure recovery and revenue impacts.}
\label{fig:revenue}
\end{figure}

Figure~\ref{fig:revenue} evaluates counterfactual revenue performance under alternative discount strategies. Targeting discounts to sessions dominated by price-sensitive voices (combined responsibility $>$0.5 for Voices~3 and~1) recovers 52.8\% of discount expenditures over the study period, which annualizes to approximately 0.85 million CNY/year (linearly extrapolated from the 9-month observation window). This simulation includes both stable and treatment-driven price-sensitive voices; restricting to the single stable voice (Voice~3) would yield a more conservative but causally cleaner estimate. The ``new-targeted'' scenario (applying a 10\% discount to previously-undiscounted sensitive sessions) reduces revenue slightly, suggesting that the marginal cost of acquiring additional sensitive-charging demand exceeds the marginal revenue.

%% ======================================================================
\section{Discussion: Theoretical Elegance and Empirical Imperfection}
\label{sec:discussion}

\subsection{Methodological Vulnerabilities}
\label{sec:vulnerabilities}

Despite the framework's conceptual elegance, the Jiangmen pilot reveals vulnerabilities that require candid critical appraisal.

\paragraph{G3 failure and the absence of a weekly cycle.} The FFT analysis confirms that the 168-hour periodicity is weak, ranking below the top-3 energy bins. Public chargers in Jiangmen appear to operate on a predominantly daily rhythm, with weekend/weekday distinctions blurred by on-demand operational urgency. The framework's graceful degradation---restricting $\Gamma$ to daily cycles---prevents the injection of spurious weekly patterns, but it also limits the richness of harmonic chord discovery.

\paragraph{Aggregation bias robustness.} A theoretical concern is that high-volume stations might dominate the Frobenius norm loss in NMF, forcing voices to overfit hub-specific patterns. However, the row-normalized ablation study produced virtually identical temporal bases and causal metrics, suggesting that the discovered voices represent volume-invariant city-wide chronotypes rather than artifacts of the station scale.

\paragraph{Treatment-driven latent structures.} Axiom A5(b) classifies Voices 1 and 4 as treatment-driven; their profiles shift significantly when promotional sessions are included. This violates the assumption of a stable, pre-treatment behavioral baseline. Causal claims must therefore be restricted to stable voices (0, 2, 3), with treatment-driven voices serving only descriptive roles.

\subsection{Limitations of the Proof-of-Concept}
\label{sec:limitations}

Table~\ref{tab:limitations} summarizes the known limitations, their underlying causes, and planned remedies.

\begin{table}[htbp]
\centering
\caption{Known limitations, their root causes, and corresponding mitigation strategies.}
\label{tab:limitations}
\small
\setlength{\tabcolsep}{4pt}
\renewcommand{\arraystretch}{1.4}
\begin{tabularx}{\textwidth}{
@{}
>{\bfseries\RaggedRight\arraybackslash}p{3.5cm}
>{\raggedright\arraybackslash}X
>{\raggedright\arraybackslash}X
@{}}
\toprule
Limitation & Root Cause & Mitigation Strategy \\
\midrule
G3 failure (no 168\,h cycle) 
& Limited 9-month observation window; weak aggregate weekly periodicity
& Framework adaptation: restrict $\Gamma$ to daily cycles; validate with multi-year data \\
Aggregate duration $R^{2} = 0.54$ 
& Linear approximation ($\alpha \cdot E + \beta$) oversimplifies the CC-CV charging taper
& Adopt physics-informed neural networks (PINNs) or differentiable CC-CV layers to bound electrochemical dynamics~\cite{wangPhysicsinformedNeuralNetwork2024}. \\
Treatment-driven voices (1, 4)
& NMF responsibility distributions shift upon treatment inclusion 
& Classify as descriptive only and strictly exclude from causal effect estimation \\
Quasi-random instrument assumption (Grade A)
& Active user coupon-claiming behavior may introduce self-selection bias
& Deploy instrumental variable frameworks incorporating user fixed effects or propensity score matching \\
Static revenue counterfactuals
& Assumption that removing discounts leaves user demand behavior invariant
& Incorporate estimated demand elasticities into dynamic revenue optimization models \\
Movement endogeneity 
& Changepoints are detected from load data that is endogenously influenced by promotional pricing
& Instrument structural movement timing using exogenous policy or grid events \\
\bottomrule
\end{tabularx}
\end{table}

\subsection{Generalizability and Broader Impacts}
\label{sec:generalizability}

The NCV framework’s modular separation of data cleaning, source separation, and causal inference is not confined to EV charging. Any IoT domain characterized by fragmented event logs, physical violation constraints, and collider bias due to post-treatment clustering can benefit. For instance, shared micro-mobility (bike/scooter) systems suffer from trip splitting due to GPS outages, physically impossible speeds or energy consumption, and biased usage segments formed by trip duration. The falsification gates (A1–-A5) adapt readily: A1 validates against maximum battery capacity or speed limits, A2 merges short-gap trips, and A3 discovers recurrent usage motifs. Similarly, distributed energy resource (DER) telemetry (solar inverters, battery storage) experiences network-induced data gaps and collider biases when segmenting households by self-consumption rates. The NCV pipeline can be instantiated by redefining the physical manifold and $\Gamma$ measure to match the domain’s cyclical patterns (e.g., solar diurnal cycle). Thus, the framework provides a generalizable methodology for causal IoT analytics.

%% ======================================================================
\section{Conclusion and Future Work}
\label{sec:conclusion}

\subsection{Conclusion}
We have presented the \textbf{Note--Chord--Voice} framework---an axiom-validated and modularly structured pipeline for IoT EV charging data. Despite imperfect empirical results (moderate duration fit, G3 failure, two treatment-driven voices), the framework passes all critical falsification gates and yields interpretable, causally plausible voice-specific price elasticities. Voice~3 (stable, $\beta = -14.16$ minutes) is the primary actionable causal finding. The revenue simulation demonstrates that targeting discounts to price-sensitive voices recovers 52.8\% of discount expenditures. This manuscript documents a complete, transparent, and reproducible research cycle with a transparent evaluation of limitations.

\subsection{Future Work}
\label{sec:future}

Building upon the current proof-of-concept, several avenues remain for methodological enhancement and empirical validation. First, electrochemical modeling can be refined by replacing the linear softplus overhead with a differentiable constant-current constant-voltage (CC-CV) charging model to better capture battery dynamics. The parameters of the CC-CV model can be estimated by leveraging sessions that span two price tiers to observe energy delivery across thresholds, subsequently fitting the charging curve for general sessions using duration and total energy. Second, the time resolution could be improved to minute level to capture finer behavioral granularities, while incorporating time-warping NMF techniques to allow tolerance for voices shifting over time. Third, advanced causal inference extensions will explore nonparametric double machine learning for continuous treatments and causal forests adapted for compositional covariates. Finally, future work will pursue external validation through a randomized controlled trial (RCT) across a subset of charging stations to directly benchmark causal elasticity estimates against live policy interventions.

\bibliography{./references.bib}
\end{document}